\documentclass[twocolumn,tightenlines,prd,floatfix,nofootinbib]{revtex4}

\usepackage{epsfig}
\usepackage{amssymb}
\usepackage{amsmath}
\usepackage{amsfonts}
\usepackage{slashed}

\newcommand{\be}{\begin{equation}}
\newcommand{\ee}{\end{equation}}

\newcommand{\nn}{\nonumber}

\newcommand{\half}{\frac{1}{2}}
\newcommand{\ztwo}{{\mathbb Z}_2}
\newcommand{\hp}{{H^\pm}}

\newcommand{\f}{\frac}

\newcommand{\eq}[1]{(\ref{#1})}

\begin{document}

\title{Associated central exclusive production of charged Higgs bosons}

\author{Rikard Enberg}
\email{rikard.enberg@physics.uu.se}
\author{Roman Pasechnik}
\email{roman.pasechnik@fysast.uu.se}

\affiliation{Department of Physics and Astronomy, Uppsala
University, Box 516, SE-751 20 Uppsala, Sweden}

\begin{abstract}
We propose central exclusive production of a
charged Higgs boson in association with a $W$ boson as a possible signature of
certain types of extended Higgs sectors.
We calculate the cross section and find
that the rate at the LHC could be large enough to
allow observation in some models with two Higgs
doublets, where the charged Higgs and at least one of the neutral
scalars can be light enough. We use the two-Higgs doublet model 
as a prototype and consider two distinct regions of parameter space, but
we also briefly discuss the prospects for the next-to-minimal supersymmetric
standard model, where the charged Higgs may very well be quite
light.
\end{abstract}

\maketitle

\section{Introduction}
The Higgs sector of the Standard Model (SM) contains a single scalar
doublet, which leads to one physical, neutral Higgs boson after electroweak
symmetry breaking. Additional Higgs bosons, and in particular a
charged Higgs boson, are predicted in many models for physics beyond
the Standard Model with extended Higgs sectors, such as the minimal
and the next-to-minimal supersymmetric Standard Models (MSSM and
NMSSM, respectively). The detection of a charged scalar would
be clear evidence of physics beyond the Standard Model. The MSSM
contains one additional Higgs doublet, but supersymmetry
places quite severe restrictions on the parameters of the
model and their relations, and enforce a Higgs sector of a special kind,
to be discussed below (the sizable loop corrections change this picture, however).
In the NMSSM, the Higgs sector contains an additional
singlet which allows for a larger variety of parameters.

It is phenomenologically interesting to consider a more minimal
addition to the SM, namely adding only one additional Higgs doublet
to the SM. In this two-Higgs doublet model (2HDM), three of the the
eight degrees of freedom give masses to the vector bosons, and five
physical Higgs bosons remain: in a CP-conserving theory these are
the CP-even neutral scalars $h^0$ and $H^0$, the CP-odd $A^0$, and
the charged Higgs bosons $H^\pm$ (see
\cite{HiggsHunters,Djouadi:2005gj} for reviews). This is a minimal
extension of the Higgs sector but it leads to a rich phenomenology
and is very useful as a laboratory for Higgs physics.

The central exclusive production (CEP) process $pp\to p + X + p$,
where $X$ stands for a centrally produced system separated from the
two very forward protons by large rapidity gaps, has been proposed
\cite{Bialas:1991wj} as an alternative way of searching for the
neutral Higgs boson (see \cite{Albrow:2010yb} for a review). The
Higgs boson is produced in the $gg\to H$ subprocess through a quark
loop. The two incoming protons survive the collision and lose only a
small fraction of their original momentum. This means that the
overall $t$-channel exchange must be a color singlet, and this
process is therefore very closely related to diffractive processes.
If the momenta of the outgoing protons are measured by forward
proton detectors placed far away from the interaction point, the
mass of the $X$ system may be reconstructed~\cite{Albrow:2000na}
with a resolution of about 2 GeV per event~\cite{FP420}. This is the proposal
of the FP420 project~\cite{FP420} which aims at placing detectors at
220~m and at 420~m away from ATLAS or CMS. The perturbative QCD
description of the CEP process began in \cite{Cudell:1995ki}, and
the calculation that is now commonly used was initiated by Khoze,
Martin, and Ryskin and collaborators in \cite{Durham}. It leads to a
cross section for the Standard Model Higgs of about 3 fb at $\sqrt{s}=14$~TeV 
for $m_H\sim 120$ GeV, but extensions of the Standard Model such as the MSSM can
yield larger cross sections. This Durham model has later been
applied for production of $\chi_c$ \cite{chic}, gluon
\cite{Cudell:2008gv} and heavy quark dijets \cite{Maciula:2010vc},
etc., and has been compared with data from the Tevatron~\cite{CDF}. We
will use this standard theoretical description in what follows.

One of the main motivations for considering central exclusive Higgs
production is that in inclusive Higgs searches, using the decay $H\to b\bar b$
is complicated due to the huge background from QCD jets. In CEP, there is a
suppression of $b\bar b$ production from
QCD events due to spin-parity conservation in the forward limit.
However, recent studies of various sources of irreducible $b\bar b$
background~\cite{Maciula:2010vc}
have revealed a potential problem at low statistics, as the
signal-to-background ratio turns out to be close to one whereas the
absolute cross section is of the order of 1~fb.
However, the overall theoretical uncertainty is
rather large (an uncertainty of about
a factor of 25 is claimed in Ref.~\cite{Dechambre:2011py})
and it is possible that the cross section is larger.
This situation makes it interesting to consider other
possible ways to probe the Higgs sector in CEP.

In this paper, we therefore propose a new potentially interesting channel,
namely central exclusive production of the charged Higgs boson in
association with a $W$ boson. This is a standard process in
inclusive searches, and we will show below that the CEP cross
section is large enough in some regions of parameter space to be
useful at the LHC. If the charged Higgs would be observed this way,
it would give important information on the properties of the Higgs
sector.

To be specific, we choose to use the 2HDM as a prototype in our
calculations. We will make one simplification. As we are here
interested in examining the feasibility of the associated production
channel, we want to consider the maximum possible cross sections.
These occur when there is an $s$-channel resonance involved,
and cross sections away from
the resonance are bound to be smaller. We will therefore concentrate
on the case when the cross section is resonantly enhanced. It has been 
shown \cite{Asakawa:2005nx} that in some regions of the parameter space
of 2HDMs,
the associated production cross section can be enhanced compared with
the MSSM by orders of magnitude.

Our results can be seen as a proof of
principle, but can be applied to more general models for physics
beyond the SM. In particular, we will briefly discuss the NMSSM as one
interesting example.

\section{The two-Higgs doublet model}

The general two-Higgs doublet model (2HDM) has two scalar doublets
$\Phi_{1,2}$ with the same hypercharge $Y=1$. Setting parameters
that break the $\ztwo$ symmetry explicitly to zero but keeping the
soft-breaking parameter $m_{12}^2$, the most general scalar
potential is given by
\begin{align}
{\cal V} &= m_{11}^2\Phi_1^\dagger\Phi_1+m_{22}^2\Phi_2^\dagger\Phi_2
-[m_{12}^2\Phi_1^\dagger\Phi_2+{\rm h.c.}]\nonumber\\
& +\half\lambda_1(\Phi_1^\dagger\Phi_1)^2
+\half\lambda_2(\Phi_2^\dagger\Phi_2)^2 \nonumber \\
&+\lambda_3(\Phi_1^\dagger\Phi_1)(\Phi_2^\dagger\Phi_2)
+\lambda_4(\Phi_1^\dagger\Phi_2)(\Phi_2^\dagger\Phi_1)
\nonumber\\
& +\half\left[\lambda_5(\Phi_1^\dagger\Phi_2)^2  + {\rm h.c.} \right]\,, \label{V}
\end{align}
where all parameters are real for CP-conserving models. Minimizing
the potential and parametrizing the doublets in terms of the
physical states, one finds relations for the masses of the Higgs
bosons in terms of $\tan\beta=v_2/v_1$, the ratio of the vacuum
expectation values of the two doublets, and the parameters of the
potential. The two CP-even scalars $h^0$ and $H^0$ mix, with mixing
angle $\alpha$. However, $\alpha$ only appears in the couplings between
Higgs bosons and gauge bosons in the combinations
$\sin(\beta-\alpha)$ and $\cos(\beta-\alpha)$. One should keep in mind that $\tan\beta$ is not
\textit{a priori} a physical parameter, and the potential \eq{V} is
in fact invariant under U(2) rotations of the doublets. In specific
models, however, a specific basis, and thus a specific value of
$\tan\beta$, is singled out as a physical parameter.

For completely general Yukawa couplings of the different Higgs
bosons,  one encounters unacceptably large flavor-changing neutral
currents mediated by Higgs exchange. Glashow and Weinberg showed \cite{Glashow:1976nt}
that these vanish if each fermion only couples to one Higgs
doublet, and one therefore usually defines four types of 2HDM,
fancifully called type I, II, III and IV, or sometimes Y and X for the last two.
The MSSM is at tree level a
type II model, where the up- and down-type fermions couple to
different doublets; however, this situation is changed somewhat by large
loop corrections. In type I, instead, all fermions couple to the
same doublet. In the following we will consider both type I and type II models.

In order for the central exclusive production mechanism to have a
cross section in the interesting range, the mass of the charged
Higgs boson must be relatively low. The experimental bounds on
$m_{H^+}$ are the strictest in the type II model, where one has,
roughly, $m_{H^+}\gtrsim 300$~GeV~\cite{Ciuchini:1997xe,Mahmoudi:2009zx}. In the
type I model, however, for
$\tan\beta\gtrsim 2$--3 there is essentially no bound beyond the model independent bound that
$m_{H^+}\gtrsim 80$~GeV~\cite{Mahmoudi:2009zx}. For this reason, we shall 
take the type I model as our principal prototype. It could also be 
interesting to consider the type X model, where $H^+$ can also be
light, see e.g.\ \cite{Aoki:2009ha} for a detailed study of its
phenomenology. The type~I results we show below
correspond to models that are allowed by all existing data.

Because it is hard to find regions in the MSSM parameter space where 
the charged Higgs is light, we shall not consider the MSSM here. In the
next-to-minimal supersymmetric Standard Model (NMSSM) on the other
hand, the charged Higgs can easily be quite light (see
e.g.~\cite{Akeroyd:2007yj,Dermisek:2008uu,Mahmoudi:2010xp}), and
central exclusive production can be interesting. Supersymmetric models 
also have a contribution to the production amplitudes from squark loops, 
which can potentially be large and positive. 

To keep the analysis simple, we will leave a detailed study of NMSSM, and 
supersymmetric models in general, for the future. Instead we will in addition 
to the type I results also show results for the 2HDM type II model, 
which is more similar to SUSY models. However, note that these type II results
are not to be taken literally, since they have light charged Higgs bosons
that are not allowed by flavor data. The reason that we still find this
interesting is that in the NMSSM the charged Higgs is allowed to be
light. The type II results should therefore be seen as an example.

\begin{figure*}[tb]
 \centerline{\epsfig{file=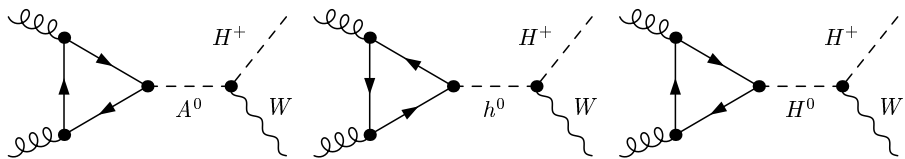, scale=0.8}}
 \caption{Subprocesses involving an $s$-channel Higgs boson.}
 \label{fig:feyn-schan}
\end{figure*}

\begin{figure}[tb]
 \centerline{\epsfig{file=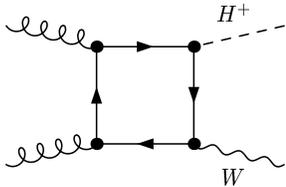, scale=0.8}}
 \caption{The box diagram.}
 \label{fig:feyn-box}
\end{figure}

\subsection{Couplings}
In the type I model, the fermions get their masses from only one of
the Higgs doublets. The dependence of the Yukawa couplings on
$\alpha$ and $\beta$ is therefore the same for the up and down type fermions. 
If the mixing
$\alpha$ is either zero or $\pi/2$, the $H^0$ or the $h^0$ does not
couple to the fermions at all. The Yukawa couplings of the three
neutral scalars to top and bottom quarks relative to the couplings
of the SM Higgs boson $\phi^0$ are in the type I model given by
\begin{align}
 \lambda_{\text{I},t}^{h^0} &=
 \lambda_{\text{I},b}^{h^0} = \frac{\cos\alpha}{\sin\beta} \label{YukIa}\\
 \lambda_{\text{I},t}^{H^0} &=
 \lambda_{\text{I},b}^{H^0} = \frac{\sin\alpha}{\sin\beta} \label{YukIb}\\
 \lambda_{\text{I},t}^{A^0} &= i\gamma_5\cot\beta \\
 \lambda_{\text{I},b}^{A^0} &= -i\gamma_5\cot\beta. \label{YukI}
\end{align}
In the type II model there is the well-known large $\tan\beta$-enhancement 
of the down-type fermion couplings. The corresponding couplings are then
\begin{align}
 \lambda_{\text{II},t}^{h^0} &= \frac{\cos\alpha}{\sin\beta} \\
 \lambda_{\text{II},b}^{h^0} &= -\frac{\sin\alpha}{\cos\beta} \\
 \lambda_{\text{II},t}^{H^0} &= \frac{\sin\alpha}{\sin\beta} \\
 \lambda_{\text{II},b}^{H^0} &= \frac{\cos\alpha}{\cos\beta} \\
 \lambda_{\text{II},t}^{A^0} &= i\gamma_5\cot\beta \\
 \lambda_{\text{II},b}^{A^0} &= i\gamma_5\tan\beta. \label{YukII}
\end{align}

The couplings in the type I model are thus the same as the couplings to
the up-type quarks in the type II model. As it is $\sin(\beta-\alpha)$
that enters the gauge boson couplings, it is useful to write these
relations in terms of $\sin(\beta-\alpha)$, $\cos(\beta-\alpha)$ and
$\tan\beta$ only,
\begin{align}
 \frac{\sin\alpha}{\sin\beta} & = \cos(\beta-\alpha) - \cot\beta \sin(\beta-\alpha)\\
 \frac{\cos\alpha}{\sin\beta} & = \sin(\beta-\alpha) + \cot\beta \cos(\beta-\alpha)\\
 \frac{\cos\alpha}{\cos\beta} & = \cos(\beta-\alpha) + \tan\beta \sin(\beta-\alpha)\\
 \frac{\sin\alpha}{\cos\beta} & =-\sin(\beta-\alpha) + \tan\beta \cos(\beta-\alpha).
\end{align}
The other important parameter for our scattering process is the
coupling of the neutral scalars to the charged Higgs and $W$. These
are the same for all types of 2HDM, and as for all Higgs--Higgs--vector
couplings, they are proportional to
$\cos(\beta-\alpha)$ for $h^0$ and to $\sin(\beta-\alpha)$ for
$H^0$.

\section{The hard subprocess}

There are four diagrams that contribute to the hard subprocess
amplitude $gg\to H^\pm W^\mp$ at the one-loop level,  when requiring
the two incoming gluons to be in a color singlet state. These are
depicted in Fig.~\ref{fig:feyn-schan} and Fig.~\ref{fig:feyn-box}.
The amplitudes for this process have been computed for inclusive
associated production by Barrientos Bendez\'u and
Kniehl~\cite{BarrientosBendezu} (see also \cite{BarrientosBendezu:2000tu,Brein:2000cv} for the MSSM results).
In the rest of this section we give results for the type I model
modified from Ref.~\cite{BarrientosBendezu}. The corresponding formulas for 
the type II model can be found in that paper.

The amplitude for $gg\to \hp W^\mp$ given by the sum of the
triangle diagrams in Fig.~\ref{fig:feyn-schan} is
\begin{align}
 V_{\lambda_W} =& \f{\sqrt{2}}{\pi} \alpha_s(\mu) G_F m_W \epsilon_{\gamma}^*(p_W)(q_1+q_2)^\gamma \nn \\
&\times \epsilon_\mu^c(q_1) \epsilon_\nu^c(q_2) \Big[ \left( q_2^\mu q_1^\nu-\f{\hat s}{2}g^{\mu\nu}\right) \Sigma(\hat s) \nn \\
& + i \epsilon^{\mu\nu\rho\sigma} q_{1\rho} q_{2\sigma}
\Pi(\hat s) \Big] , \label{Vhard}
\end{align}
where $\alpha_s(\mu)$ is the strong coupling, $\hat s=M_{HW}^2$ is
the invariant mass squared of the $\hp W^{\mp}$ pair, $\mu$ is the
renormalization scale, $\epsilon_{\gamma}^*$ is the polarization
vector of the $W$ boson with momentum $p_W$ and helicity
$\lambda_W$, and $\epsilon_{\mu,\nu}^c$ are the polarization vectors
of the gluons with momenta $q_{1,2}$. These are summed over the
color index $c$. The functions $\Sigma$ and $\Pi$ come from the loop
integration and correspond to $h^0$ and $H^0$ exchange ($\Sigma$) and
$A^0$ exchange ($\Pi$) in the $s$-channel. They are given by
\begin{align}
 \Sigma(\hat s) &= \sum_{q=t,b} {\cal S}(\hat s) S\left(\f{\hat s+i\epsilon}{4m_q^2}\right) \\
 \Pi(\hat s) &= \sum_{q=t,b} {\cal P}_q(\hat s) P\left(\f{\hat s+i\epsilon}{4m_q^2}\right) ,
\end{align}
where the functions
\begin{align}
 S(r) &= \f{1}{r}\left[ 1-\left(1-\f{1}{r}\right) \text{arcsinh}^2\sqrt{-r} \right] \\
 P(r) &= -\f{1}{r} \text{arcsinh}^2\sqrt{-r}
\end{align}
must be continued analytically for three regions in $r$, such that
for $r\le 0, 0<r\le 1$, or $r>1$ one must use
$\text{arcsinh}\sqrt{-r}, -i \arcsin\sqrt{r}$, or
$\text{arccosh}\sqrt{r} -i\pi/2$. The functions ${\cal S}$ and
${\cal P}$ contain the propagators and relative couplings and are
defined as
\begin{align}
 {\cal S}(\hat s) &= \f{1}{\sin\beta} \left( \f{\cos\alpha \, \cos(\alpha-\beta)}{\hat s - m_{h^0}^2 + i m_{h^0} \Gamma_{h^0}} \right. \nn \\
&\quad\quad\quad \left. + \f{\sin\alpha \, \sin(\alpha-\beta)}{\hat s - m_{H^0}^2 + i m_{H^0} \Gamma_{H^0}} \right) \\
{\cal P}_t(\hat s) &= \f{\cot\beta}{\hat s - m_{A^0}^2 + i m_{A^0}
\Gamma_{A^0}},
\end{align}
and ${\cal P}_b(\hat s) = -{\cal P}_t(\hat s)$.
We have modified the ${\cal S,P}$ functions given in~\cite{BarrientosBendezu}
with the appropriate Yukawa couplings for type I. Thus the $t$ and $b$ functions are
identical in our case. However, the contribution from $b$ is
negligible, since the $S$ and $P$ functions tend to zero for $r\to\infty$,
and there is no $\tan\beta$ enhancement in type I.

We do not list the complicated expressions for the box diagrams,
schematically shown in Fig.~\ref{fig:feyn-box}.

As discussed above, if the mass relations are such that one of the
intermediate Higgs bosons $h^0$, $H^0$ or $A^0$ is close in mass to
the $H^\pm  W^\mp$ system, the three resonant triangle diagrams in
Fig.~\ref{fig:feyn-schan} will completely dominate the amplitude. We
have checked this fact by calculating the hard $gg\to \hp W^{\mp}$
subprocess cross section at the $h^0$ and $H^0$ resonances in two
ways:\ exactly, with triangle and box diagrams included, and keeping
triangles only. These calculations were performed using FeynArts and
FormCalc \cite{FAFC}. The relative numerical difference between
these two cross sections is extremely small, on the order of
$10^{-6}$, meaning that the interference between triangles and boxes
at the Higgs resonance is totally negligible. In this paper we will
concentrate on scenarios that yield the largest possible cross
sections, and we will therefore neglect the box diagrams.

In inclusive associated production, all three triangle diagrams
contribute to the amplitude. This is not the case for central
exclusive production, which occurs in the forward limit. As we will
show below, the amplitude with an $s$-channel $A^0$ boson vanishes
in this limit due to its CP-odd nature. We therefore only need to
consider the amplitudes with exchange of $h^0$ and $H^0$.

\section{Central exclusive production}

\begin{figure}[tb]
 \centerline{\epsfig{file=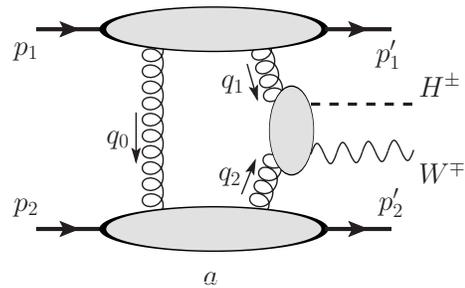,width=6.0cm}}
 \caption{The central exclusive $H^{\pm}W^{\mp}$ pair production.
 Typical contributions to the hard subprocess scattering
 amplitude $gg\to H^{\pm}W^{\mp}$ are shown
 in Figs.~\ref{fig:feyn-schan} and \ref{fig:feyn-box}.}
 \label{fig:HW-CEP}
\end{figure}

We follow the QCD mechanism for central exclusive
production, initially developed by Khoze, Martin and Ryskin
(KMR) in Refs.~\cite{Durham}. A schematic diagram
for central exclusive associated $H^{\pm}W^{\mp}$ pair production in
proton-proton scattering $pp\to pH^{\pm}W^{\mp}p$ is shown in
Fig.~\ref{fig:HW-CEP}.

The momenta of the intermediate gluons are given by Sudakov
decompositions in terms of the incoming proton momenta $p_{1,2}$
\begin{eqnarray}\nonumber
&&q_1=x_1p_1+q_{1\perp},\quad q_2=x_2p_2+q_{2\perp},\\
&&q_0=x'p_1-x'p_2+q_{0\perp}\simeq q_{0\perp},\quad x'\ll
x_{1,2},\label{moms}
\end{eqnarray}
such that $q_{\perp}^2\simeq -|{\bf q}|^2$. Here, and below, we write
transverse 2-momenta in boldface. In the
forward scattering limit, we have
\begin{eqnarray}\nonumber
&&t_{1,2}=(p_{1,2}-p'_{1,2})^2={p'}^2_{1/2\perp}\to0,\\
&&q_{0\perp}\simeq-q_{1\perp}\simeq q_{2\perp}\,. \label{forward}
\end{eqnarray}

According to the KMR approach, we write the
amplitude of this process, which in the diffractive limit is dominated
by its imaginary part, as
\begin{eqnarray} \label{ampl}
{\cal M}_{\lambda_W}&\simeq&is\frac{\pi^2}{2}\frac{1}{N_c^2-1} \int
d^2 {\bf q}_0V_{\lambda_W}\frac{f_g(q_0,q_1)f_g(q_0,q_2)} {{\bf
q}_0^2\,{\bf q}_1^2\,{\bf q}_2^2}\,,\nonumber
\end{eqnarray}
where $\lambda_W$ is the helicity of produced $W^{\pm}$ boson,
$f_g(r_1,r_2)$ is the off-diagonal unintegrated gluon distribution
function (UGDF), which is dependent on the longitudinal and transverse
components of both gluons $r_1$ and $r_2$ emitted from the proton
line. The $gg\to \hp W^{\mp}$ hard subprocess amplitude
$V_{\lambda_W}$ in given by Eq.~(\ref{Vhard}). The diffractive
amplitude (\ref{ampl}) is averaged over the color indices and over
the two transverse polarizations of the incoming gluons.

The bare amplitude above is subject to absorption corrections which
depend on the collision energy and the typical proton transverse
momenta. In the original KMR calculations the bare production cross
section is simply multiplied by a gap survival factor, which is
estimated to be $\hat{S}^2\simeq 0.015$ at the LHC energy \cite{SF}.

\subsection{Unintegrated gluon distributions}

The coupling of the gluons to the proton is described in terms of the off-diagonal
unintegrated gluon distribution functions (UGDFs)
$f_g(q_0,q_{1,2})=f^{\mathrm{off}}_g(x',x_{1,2},{\bf q}_{1,2}^2,{\bf
q}_0^2,\mu_F^2;t_{1,2})$ at the factorization scale
$\mu_F \sim M_{HW}\gg |{\bf q}_0|$. In
the forward (\ref{forward}) and asymmetric limit of
small $x'\ll x_{1,2}$, where $x'$ is the longitudinal momentum fraction of
the screening gluon, the off-diagonal UGDF is
written as a skewedness factor $R_g$ multiplying the
diagonal UGDF, which describes the coupling of gluons with momentum
fractions $x_{1,2}$ to the proton (see Refs.\ \cite{Kimber:2001sc,MR} for details).
The skewedness parameter $R_g\simeq 1.2-1.3$ is expected to be roughly constant at
LHC energies and gives only a small contribution to the
overall normalization uncertainty.

In the kinematics considered here, the unintegrated gluon density can
be written in terms of the conventional gluon distribution
$g(x,{\bf q}^2)$ as~\cite{MR}
\begin{equation}\label{ugdfkmr}
f_g(x,{\bf q}^2,\mu_F^2)=\frac{\partial}{\partial\ln {\bf q}^2}
\big[xg(x,{\bf q}^2)\sqrt{T_g({\bf q}^2,\mu_F^2)}\big] \; ,
\end{equation}
where $T_g$ is the Sudakov form factor which suppresses real
emissions during the evolution, so that the
rapidity gaps are not populated by gluons. It is given by
\begin{eqnarray}\nonumber
T_g({\bf q}^2,\mu_F^2)&=&{\rm exp} \bigg(-\int_{{\bf q}^2}^{\mu_F^2}
\frac{d {\bf k}^2}{{\bf k}^2}\frac{\alpha_s({\bf k}^2)}{2\pi}\times\\
&&\int_{0}^{1-\Delta} \!\bigg[ z P_{gg}(z) + \sum_{q} P_{qg}(z)
\bigg]dz \!\bigg),\label{Sudak}
\end{eqnarray}
where $\Delta$ in the upper limit is taken to be \cite{Coughlin:2009tr}
\begin{equation}\label{delta}
\Delta=\frac{|{\bf k}|}{|{\bf k}|+M_{HW}} \; .
\end{equation}

\subsection{CEP as a spin-parity analyzer}

Due to its CP-odd nature, the central exclusive $A^0$
production is suppressed in the forward limit due to what has become known as the
$J_z=0$ selection rule \cite{Durham}. To demonstrate this, let us calculate explicitly
the hard subprocess part $V_{\lambda_W}$
(\ref{Vhard}) describing the scattering of two basically on-shell
gluons into an $\hp W^{\mp}$ pair. Summing over colors and
polarizations of the gluons, we have
\begin{align}
 V_{\lambda_W} =& (N_c^2-1)\f{\sqrt{2}}{\pi} \alpha_s(\mu) G_F m_W \epsilon_{\gamma}^*(p_W)(q_1+q_2)^\gamma \nn \\
&\times n_\mu^{-}n_\nu^{+} \Big[ \left( q_2^\mu q_1^\nu-\f{\hat s}{2}g^{\mu\nu}\right) \Sigma(\hat s) \nn \\
& + i \epsilon^{\mu\nu\rho\sigma} q_{1\rho} q_{2\sigma}
\Pi(\hat s) \Big]\,,\quad
n_{\mu}^{\mp}=\frac{p_{1,2}^{\mu}}{E_{p,\mathrm{cms}}},
\label{Vhard-1}
\end{align}
where by a convention we adopt the lightcone vectors $n_{\mu}^{\pm}$ as
transverse gluon polarization vectors, and where
$E_{p,\mathrm{cms}}=\sqrt{s}/2$. Momentum conservation and gauge invariance imply that
\begin{eqnarray}\nonumber
\hat s = x_1x_2s\simeq 2(q_1q_2),\;\;
V_{\lambda_W}=n_\mu^{-}n_\nu^{+}V^{\mu\nu}=\frac{4}{\hat s}
q_{1\perp}^{\mu}q_{2\perp}^{\nu}V_{\mu\nu}.
\end{eqnarray}
A straightforward calculation leads to
\begin{align}
 V_{\lambda_W} =& -(N_c^2-1)\f{2\sqrt{2}}{\pi}\, \alpha_s(\mu) G_F m_W (\epsilon^*(p_W)\cdot p_H) \nn \\
&\times \Big[(q_{1\perp}q_{2\perp})\Sigma(\hat s)+i
(q_2^xq_1^y-q_2^yq_1^x) \Pi(\hat s) \Big]\,, \label{Vhard-fin}
\end{align}
from which it is obvious that in the forward limit
given by Eq.~(\ref{forward}), the coefficient in front of $\Pi(\hat
s)$ disappears, so the contribution of $A^0$ to central
exclusive $\hp W^{\mp}$ pair production vanishes, and only the
$h^0,\,H^0$ contributions to $\Sigma(\hat s)$ survive.

\subsection{$\hp W^{\mp}$ CEP cross section in the narrow-width approximation}

As we consider resonance production, we use the narrow-width
approximation in our calculation of the cross section, and therefore
need the production cross section of $h^0$ and $H^0$. In the type I
2HDM, there is no large-$\tan\beta$ enhancement of Yukawa couplings
to $b$-quarks; thus the contribution from $b$-quark loops to the
$gg\to h^0,\,H^0$ process is negligible. The only difference between
the CEP cross section for the Standard Model Higgs boson $H$ and for
the 2HDM Higgs bosons $h^0,\,H^0$ is then through the Yukawa
couplings defined in Eqs.~(\ref{YukIa}, \ref{YukIb}). The central exclusive
associated $\hp W^{\mp}$ production in the narrow-width
approximation is then given by the contribution from the relevant
resonance, either $h^0$ or $H^0$, as
\begin{align}
\sigma^{\mathrm{CEP}}_{\mathrm{HW}} \simeq
\begin{cases}
\displaystyle\sigma^{\mathrm{CEP}}_{h_{\mathrm{SM}}}(m_{h^0})(\lambda_{\text{I},t}^{h^0})^2 \mathrm{BR}(h^0\to
\hp W^{\mp}) \\
\displaystyle\sigma^{\mathrm{CEP}}_{h_{\mathrm{SM}}}(m_{H^0})(\lambda_{\text{I},t}^{H^0})^2 \mathrm{BR}(H^0\to
\hp W^{\mp})
\end{cases}
\label{WH-CEP}
\end{align}
where $\sigma^{\mathrm{CEP}}_{h_{\mathrm{SM}}}(m_h)$ is the Standard
Model Higgs boson CEP cross section calculated at a given Higgs mass
$m_h$.

In the type II model, on the other hand, the contribution from $b$-quarks can be significant.
Since we are working in the narrow-width approximation, the contribution from $b$-quark loops
cannot be added coherently on the amplitude level. We therefore add this contribution 
on the cross section level, ignoring the interference terms. 
We estimate from the sizes of the couplings that this will 
give an error of less than 20\%. A second approximation is that the cross section for the
Standard model Higgs is computed for $t$-quark loops only, and we now want to use this
result for $b$-quark loops. Referring to Eq.~(\ref{Vhard}) and the slow variation of the 
function $S(r)$, we estimate that the error we make here is less than 5\%. Within these
approximations, the central exclusive cross section for type II is then given, for $h^0$ or $H^0$, by
\begin{align}
\sigma^{\mathrm{CEP}}_{\mathrm{HW}} \simeq
\begin{cases}
\displaystyle\sigma^{\mathrm{CEP}}_{h_{\mathrm{SM}}}(m_{h^0})
\left[(\lambda_{\text{II},t}^{h^0})^2+(\lambda_{\text{II},b}^{h^0})^2\right] \\
\hfill \times \mathrm{BR}(h^0\to \hp W^{\mp}) \\
\displaystyle\sigma^{\mathrm{CEP}}_{h_{\mathrm{SM}}}(m_{H^0})
\left[(\lambda_{\text{II},t}^{H^0})^2+(\lambda_{\text{II},b}^{H^0})^2\right] \\
\hfill \times \mathrm{BR}(H^0\to \hp W^{\mp}) 
\end{cases}
\label{WH-CEPII}
\end{align}

We have checked that the narrow-width approximation works in the type I case by also
computing the full $2\to 4$ cross section $\sigma(pp\to p H^+W^-p)$
using the hard subprocess formulas in Eq.~(\ref{Vhard}) and comparing with
the results from Eq.~(\ref{WH-CEP}) for some parameter points.

\begin{figure}[tb]
 \centerline{\epsfig{file=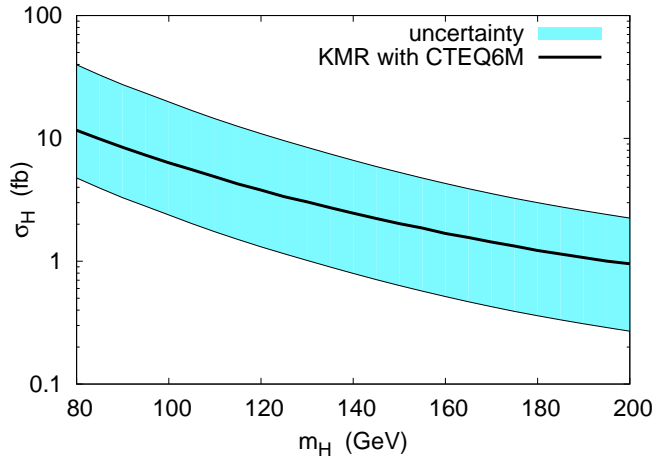,width=\columnwidth}}
 \caption{Cross section for exclusive SM Higgs boson production
 at the LHC at 14 TeV as a function of the Higgs boson mass. The thick line is
 the KMR result obtained using the CTEQ6M pdf. The thin lines
 illustrate the theoretical uncertainty (about a factor of ten); see the description in the text.}
 \label{fig:SM-CEP}
\end{figure}

We will consider the cross section
$\sigma^{\mathrm{CEP}}_{h_{\mathrm{SM}}}(m_h)$ calculated in the KMR
model. The main sources of uncertainties are the unintegrated
generalized gluon distribution, the gap survival probability factor,
and the scale choice in the Sudakov form factor. In
Fig.~\ref{fig:SM-CEP} we display the cross section for a SM Higgs
boson as a function of the Higgs mass together with an uncertainty
band. The upper line is given by the largest KMR result, as quoted
in \cite{Albrow:2010yb}, which is obtained using the CTEQ6L parton
distribution. The lower line is given by the smallest KMR result, as
quoted in \cite{Dechambre:2011py}, which is obtained by using a
modified scale choice in the Sudakov form factor, as prescribed in
Ref.\ \cite{Coughlin:2009tr}. The uncertainty is roughly a factor
ten.

The central line is the result we use in the present study. This
line is given by the KMR result, quoted in \cite{Albrow:2010yb},
obtained by using the CTEQ6M parton distribution.

For a given point in the 2HDM parameter space we
can thus easily obtain $\sigma^{\mathrm{CEP}}_{\mathrm{HW}}$  using
Eqs.~(\ref{WH-CEP}) or (\ref{WH-CEPII}).

\section{Parameter scans}

\begin{figure*}[t]
\begin{center}
\begin{tabular}{cc}
\epsfig{file=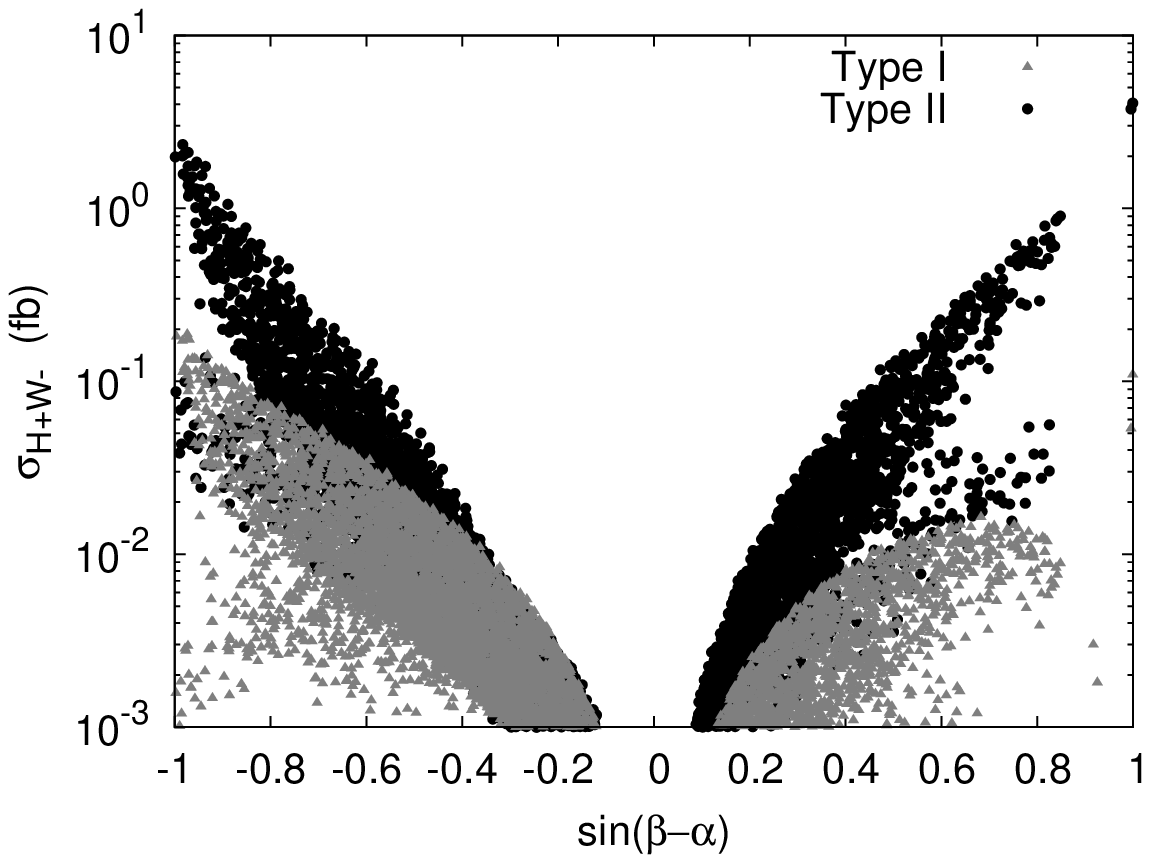,width=\columnwidth} & \epsfig{file=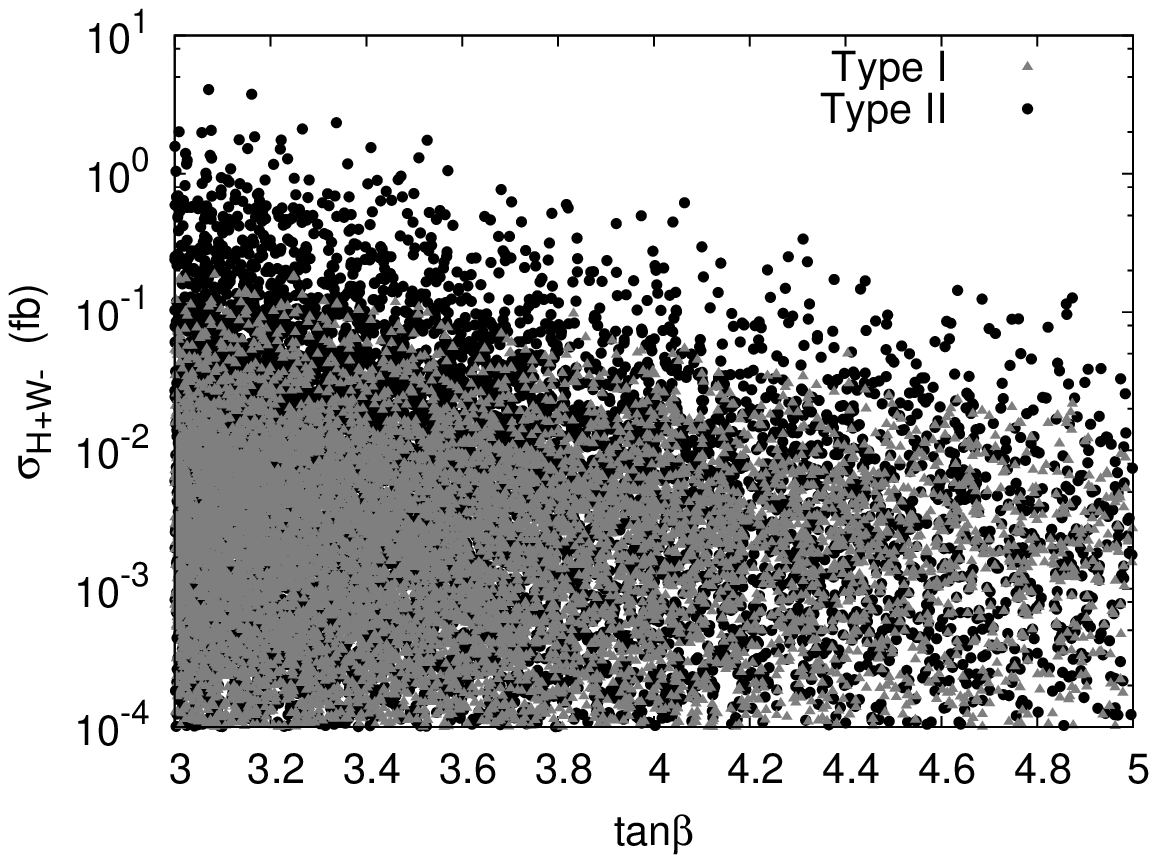,width=\columnwidth}\\
~ ~ ~ ~ ~ \textbf{(a)} & ~ ~ ~ ~ ~ ~ \textbf{(b)}\\[1mm]
\epsfig{file=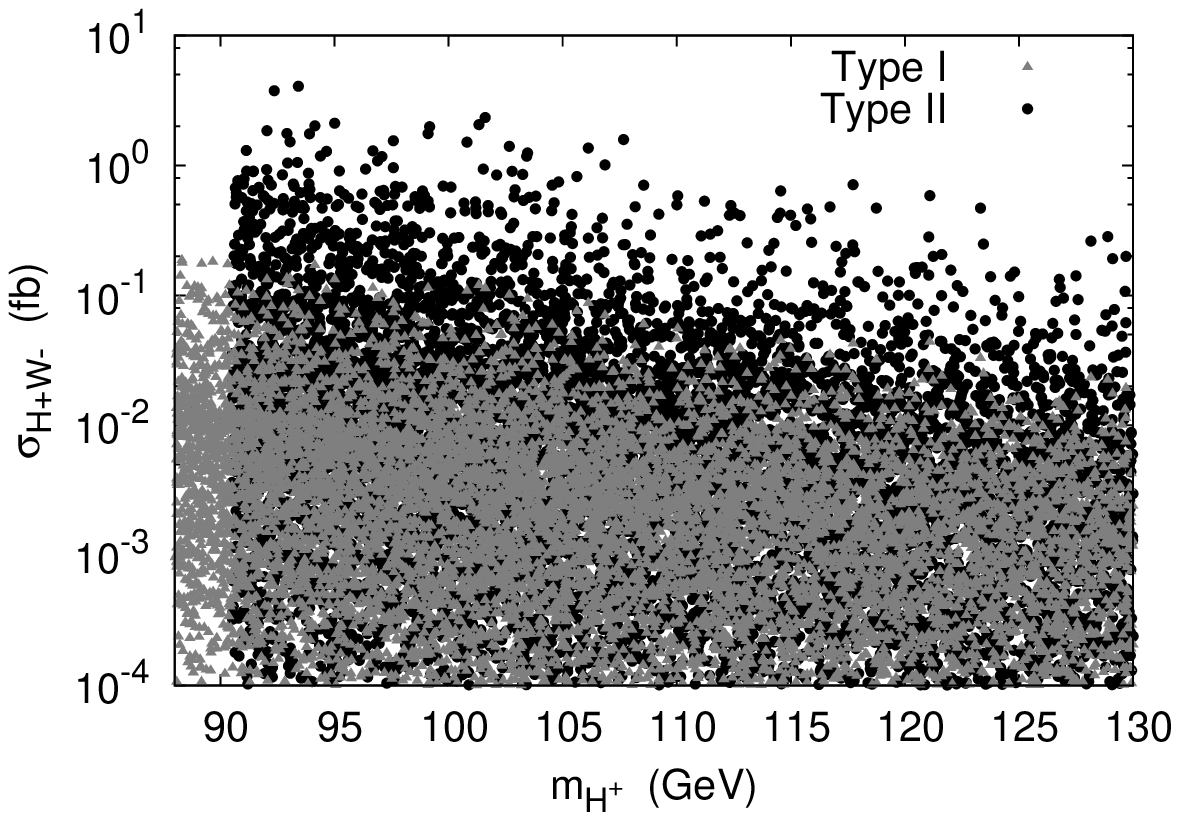,width=\columnwidth} & \epsfig{file=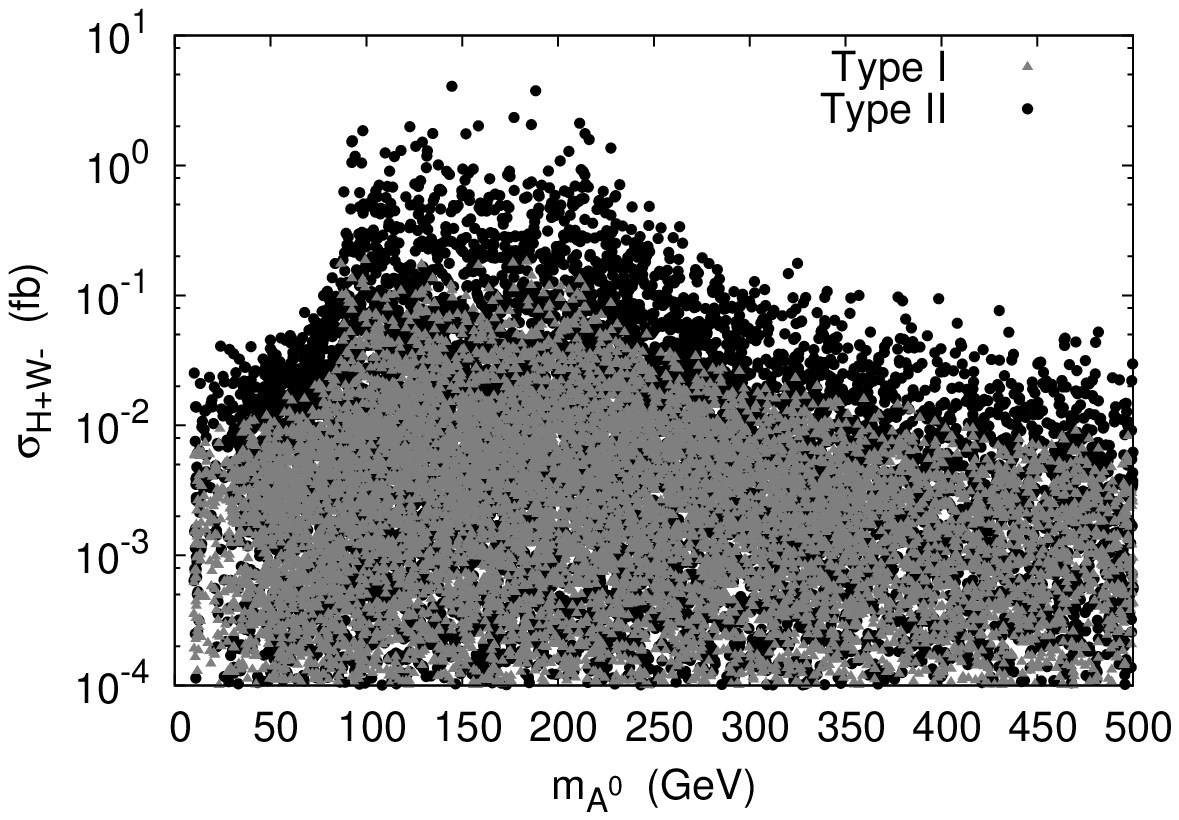,width=\columnwidth}\\
~ ~ ~ ~ ~ \textbf{(c)} & ~ ~ ~ ~ ~ ~ \textbf{(d)}
\end{tabular}
\end{center}
 \caption{Cross sections for central exclusive associated $H^+W^-$ production
 at $\sqrt{s}=14$ TeV at the LHC for points in the parameter scan.}
 \label{fig:sigma-CEP}
\end{figure*}

In order to examine if the cross sections for associated CEP can be
large enough to be observed at LHC, we perform a simple scan over
the parameter space of the model. For this purpose, we use the 2HDMC
code~\cite{Eriksson:2009ws} to compute the relevant parameters and
masses of the considered models. 2HDMC is a public code that
computes the masses and couplings of a general 2HDM from a specified
set of input parameters of the potential, and also features a
completely general Yukawa sector, which can also be restricted to
type I, II, III or IV Yukawa sectors. It further includes both
theoretical and experimental checks on the obtained model, and
features a link to HiggsBounds~\cite{Bechtle:2008jh} which allows
further checks against collider data.

We scan over the parameter space of type I and II two-Higgs doublet models
in the physical basis, defined as the parameter basis where one replaces
potential parameters $\lambda_i$ with the physical Higgs boson masses
as input parameters. The parameters are then $m_{h^0}$, $m_{H^0}$,
$m_{A^0}$, $m_{H^+}$, $m_{12}^2$, $\tan\beta$ and $\sin(\beta-\alpha)$.
We choose points with $m_{H^0}=m_{H^+}+m_W$ to
maximize the cross section in the $H^0$-exchange channel, and following
the convention we are using, we impose $m_{h^0}<m_{H^0}$.

\begin{table}
\begin{tabular}[t]{|l|c|c|c|c|c|c|}
\hline
× & $m_{h^0}$ & $m_{A^0}$ & $m_{H^+}$ & $m_{12}^2$ & $\tan\beta$ & $\sin(\beta-\alpha)$\\
\hline
lower & 115  & 10 & 88 & 500 & 3 & $-1$\\
upper & 160  & 500 & 130 & 1500 & 5 & 1\\
\hline
\end{tabular}
\caption{Parameter ranges in scan. Dimensionful parameters are given
in GeV.\label{table}}
\end{table}

In the scan, we generate parameter points, choosing all parameters
from a flat distribution within the limits shown in
Table~\ref{table}. For each generated point we check positivity of
the Higgs potential, unitarity at tree level and perturbativity. We
further check the electroweak precision constraints on the oblique
parameters $S$, $T$ and $U$~\cite{Peskin:1990zt}, the constraint on
$g-2$ of the muon~\cite{gmuon}, and that the obtained Higgs masses
are not ruled out by collider constraints. These checks are all
performed using 2HDMC and HiggsBounds, and are applied to both the 
type I and type II models.

Additionally, one may take constraints from flavor physics into
account. These constraints are not included in 2HDMC, but a detailed
analysis has been published in Ref.~\cite{Mahmoudi:2009zx}. The main
flavor physics constraint for our purposes is that $\tan\beta >3$,
which is included in the choice of parameter limits. Note that, while 
all points we show for type I satisfy all constraints, as
pointed out above, the type II model is ruled out by the flavor 
constraints and is shown instead as an example of a different type
of Higgs sector, which can be relevant for extended versions of 
supersymmetry.

We generate $10^4$ points that pass the constraints. For each such
point we compute the total central exclusive cross section using
Eqs.\ (\ref{WH-CEP}) or (\ref{WH-CEPII}). The results are shown in
Fig.~\ref{fig:sigma-CEP}, where we show scatter plots of the cross
section versus $\sin(\beta-\alpha)$, $\tan\beta$, $m_{H^+}$, and
$m_{A^0}$ for both type I and type II. 

Fig.~\ref{fig:sigma-CEP}(a) shows
that for $|\sin(\beta-\alpha)|\gtrsim 0.6$, the cross sections 
in the type II model are larger than 0.1~fb, and for  
$\sin(\beta-\alpha)\lesssim -0.9$, they exceed 1~fb. The cross sections
in the type I model do not vary as sharply with $\sin(\beta-\alpha)$,
and are therefore smaller by factors of a few to a factor of ten.

The large cross sections occur in the region where the $h^0-H^0$ mixing
becomes small so that the coupling of the $H^0$ to the fermions is
maximal and that of $h^0$ is minimal. In these regions, the cross
sections can thus be large enough to perhaps allow detection at LHC
(note that the proposed cross sections for CEP of the SM Higgs boson
are in some estimations of the order of 0.1~fb). The dependence on
$\tan\beta$ shown in Fig.~\ref{fig:sigma-CEP}(b) is not as strong,
but smaller values closer to the lower limit yield larger cross
sections. The dependence on the charged Higgs mass shown in
Fig.~\ref{fig:sigma-CEP}(c) is not strong within the limits, and it
might be possible to consider larger values than we have done here,
where we specialize to light $H^\pm$. The CP-odd Higgs mass, shown
in Fig.~\ref{fig:sigma-CEP}(d), on the other hand, is preferred to
be between 100 GeV and 250 GeV. For
higher masses, the theoretical constraints on the Higgs potential
become important and reduce the available parameter space. The drop
in cross sections below 100 GeV is reflected in
Fig.~\ref{fig:sigma-CEP}(a), where the structure at $\sigma\sim
10^{-2}$~fb at small $\sin(\beta-\alpha)$ corresponds to lower
$m_{A^0}$.

To summarize the parameter scans, there are regions of parameter
space of our selected prototype model where the cross sections are
large enough to conceivably allow detection at the LHC.

\section{Detection prospects, backgrounds}

Experimentally, CEP will be searched for in high luminosity running
at LHC with the help of forward proton detectors. We only consider 
the LHC at 14~TeV, since the luminosity at 7~TeV will not be large 
enough. One might expect
that detection at high luminosity would be complicated by pile-up,
but it has been shown that, at least for $H\to b\bar b$, this
problem can be overcome through careful cuts and vertex
reconstruction \cite{Cox:2007sw}. Pile-up can also be reduced by
timing measurements of the forward protons \cite{Albrow:2000na}.

The mass reconstruction of the central system is effective
regardless of the decay channels of the central system. Since only
forward, small angle scattering is considered, the outgoing protons
will have small transverse momenta, so that also the centrally
produced system has a small transverse momentum. The charged Higgs
and the $W$ boson will therefore be more or less back-to-back. Since
we are interested in higher masses of the central system than the
canonical 120~GeV, the suggested forward detectors at 220~m in
addition to the ones at 420~m would increase the acceptance
\cite{Heinemeyer:2007tu}.

The main decay channels of a light charged Higgs boson (light
meaning lighter than the top quark) in the type I and II models are $H^+\to
\tau^+\nu$ and $H^+\to c\bar s$. There are therefore several possible
scenarios. If both the $H^+$ and the $W$ decay leptonically, there
will be a large amount of missing energy due to the neutrinos,
together with a $\tau$ and a lepton. If they both decay
hadronically, there will be four jets. If $m_{H^+}$ is close to
$m_W$, special care may be needed to distinguish the associated production
process from a SM Higgs that decays into $W^+W^-$ \cite{WW}.

If one should consider the NMSSM, the decay channel $H^+\to W^+ A_1$ can be prominent, since 
$A_1$ is light in many regions of parameter space. Here $A_1$ 
is the lightest of the CP-odd neutral Higgs bosons.  (In a few points in our 
parameter space the decay $H^+\to W^+ A^0$ is also significant, but these points
have low cross sections.)

There is much less background to the associated production signal
than to the SM $H\to b\bar b$ signal, where there is an irreducible
$b\bar b$ background. Backgrounds that may need to be considered
include the $gg\to W+\text{jets}$, $gg\to WW$, $\gamma\gamma\to WW$
and $\gamma\gamma\to W\ell\nu$ processes considered in \cite{WW}. Thus,
even if the associated production cross section is smaller than the 
SM Higgs cross section, the significance for $H^+W^-$ could be 
as large or even larger.

\section{Conclusions and outlook}
In this paper we have proposed a novel central exclusive production
mechanism for charged Higgs bosons in association with a $W$ boson.
We have computed the cross section for this channel in prototype
two-Higgs doublet models with light charged Higgs bosons. 
We have performed a limited parameter scan over the
parameters of the model and have found that in some parts of
parameter space, where the mixing of the CP-even Higgs bosons is
small, the cross sections can be large enough to allow detection at
LHC.

In the NMSSM, the charged Higgs boson is allowed to be rather light, 
and as our calculations using the type II 2HDM show, the associated CEP
cross section can be almost as large as the SM Higgs production. 
It would therefore be very interesting to investigate this process in 
the NMSSM.

\section*{Acknowledgments}

We thank Gunnar Ingelman, Oscar St{\aa}l and Antoni Szczurek for useful discussions.
This work was supported by the Swedish Research Council under
Contract No.\ 2007-4071 and by the Carl Trygger Foundation.



\end{document}